\documentclass[twocolumn,preprintnumbers,amsmath,amssymb,prb]{revtex4-1}

\usepackage{graphicx}
\usepackage{dcolumn}
\usepackage{subfigure}
\usepackage{bm}

\begin{document}


\title{Electronic structure of the high and low pressure polymorphs of MgSiN$_{2}$}

\author{M. R{\aa}sander}\email{m.rasander@imperial.ac.uk}
\affiliation{%
Department of Materials, Imperial College London, SW7 2AZ, London, United Kingdom
}%
\author{M. A. Moram}
\affiliation{%
Department of Materials, Imperial College London, SW7 2AZ, London, United Kingdom
}%
\date{\today}

\begin{abstract}
We have performed density functional calculations on the group II-IV nitride MgSiN$_{2}$. At a pressure of about 20~GPa the ground state wurtzite derived MgSiN$_{2}$ structure (LP-MgSiN$_{2}$) transforms into a rock-salt derived structure (HP-MgSiN$_{2}$) in agreement with previous theoretical and experimental studies. Both phases are wide band gap semiconductors with indirect band gaps at equilibrium of 5.58~eV (LP-MgSiN$_{2}$) and 5.87~eV (HP-MgSiN$_{2}$), respectively. As the pressure increases, the band gaps become larger for both phases, however, the band gap in LP-MgSiN$_{2}$ increases faster than the gap in HP-MgSiN$_{2}$ and with a high enough pressure the band gap in LP-MgSiN$_{2}$ becomes larger than the band gap in HP-MgSiN$_{2}$.
\end{abstract}

\maketitle
\section{Introduction}

The Group II-IV nitride semiconductors can be derived from Group III nitrides by substituting pairs of group III (Al, Ga or In) atoms for a single group II (Be, Mg, Ca, or Zn) atom and a single group IV (C, Si, Ge or Sn) atom. Depending on which combination of Group II and Group IV elements that is chosen the II-IV nitrides have possible applications ranging from solar cells to ultra violet light emitting diodes.\cite{Punya2011} In the case of MgSiN$_{2}$ the substitution leads to an ordered wurtzite-like orthorhombic (space group Pna2$_{1}$, no. 33) crystal structure.\cite{David1970,Bruls1} MgSiN$_{2}$ has been found to possess many favourable physical properties, such as a high thermal conductivity, thermal stability and high hardness. It has also been found to have a large band gap, comparable to AlN.\cite{Quirk,deBoer2015} The band gap is indirect\cite{Quirk,deBoer2015} and has been measured to be 5.7(2)~eV\cite{deBoer2015} in agreement with theoretical calculations.\cite{Quirk,deBoer2015} This value is about 1~eV larger than an earlier value of 4.8~eV obtained by Gaido {\it et al.}\cite{Gaido1974} using diffuse reflectance spectra. However, it is now clear from theory and experiment that the band gap of MgSiN$_{2}$ is comparable in size to the band gap in wurtzite AlN.
\par
It was predicted theoretically\cite{Fang2,Romer2009} and later confirmed experimentally\cite{Andrade2011} that at the rather moderate pressure of about 25~GPa, MgSiN$_{2}$ transforms from the low pressure (LP) orthorhombic phase, where all atoms are in a wurtzite-like surrounding, into a rhombohedral caswellsilverite (NaCrS$_{2}$) structure type (space group R$\bar{3}$m, no. 166), where all atoms are found in a rock-salt like octahedral surrounding.\cite{Romer2009,Andrade2011} Furthermore, the high pressure (HP) phase was found to be meta-stable at ambient conditions.\cite{Andrade2011} At present, there are a number of theoretical studies on the phase transition from orthorhombic to rhombohedral MgSiN$_{2}$,\cite{Fang2,Romer2009,Arab2015} however, an investigation of the electronic structure of the HP-phase in relation to LP-MgSiN$_{2}$ has so far not been performed. The aim of the present study is, therefore, to investigate the electronic structure of HP-MgSiN$_{2}$ in relation to LP-MgSiN$_{2}$. In addition, we also provide results regarding the phase stability of the high and low pressure phases. For this reason we have performed density functional calculations using a range of traditional density functional approximations as well as the hybrid density functional approximation of Heyd, Scuseria and Ernzerhof (HSE)\cite{Heyd2003,Heyd2006}. For orthorhombic MgSiN$_{2}$, it has been shown that traditional exchange-correlation approximations give band gaps in the range of 4.1-4.6~eV.\cite{Huang,Fang1,Fang2}
 These values are in rather good agreement with the experimental value of Gaido {\it et al.},\cite{Gaido1974} however, they are too small in comparison to the recent observation of 5.7(2)~eV by de Boer {\it et al.}\cite{deBoer2015} As mentioned previously, recent calculations using hybrid density functional theory,\cite{Quirk} or the modified Becke-Johnson approximation,\cite{deBoer2015} provide larger band gaps in very good agreement with the measured band gap of de Boer {\it et al.} for the LP-phase of MgSiN$_{2}$. This shows the importance of using approximations that are able to obtain reliable band structures and therefore to go beyond using standard local and semilocal density functional approximations for these types of materials.

\begin{figure}[b]
\includegraphics[width=6.5cm]{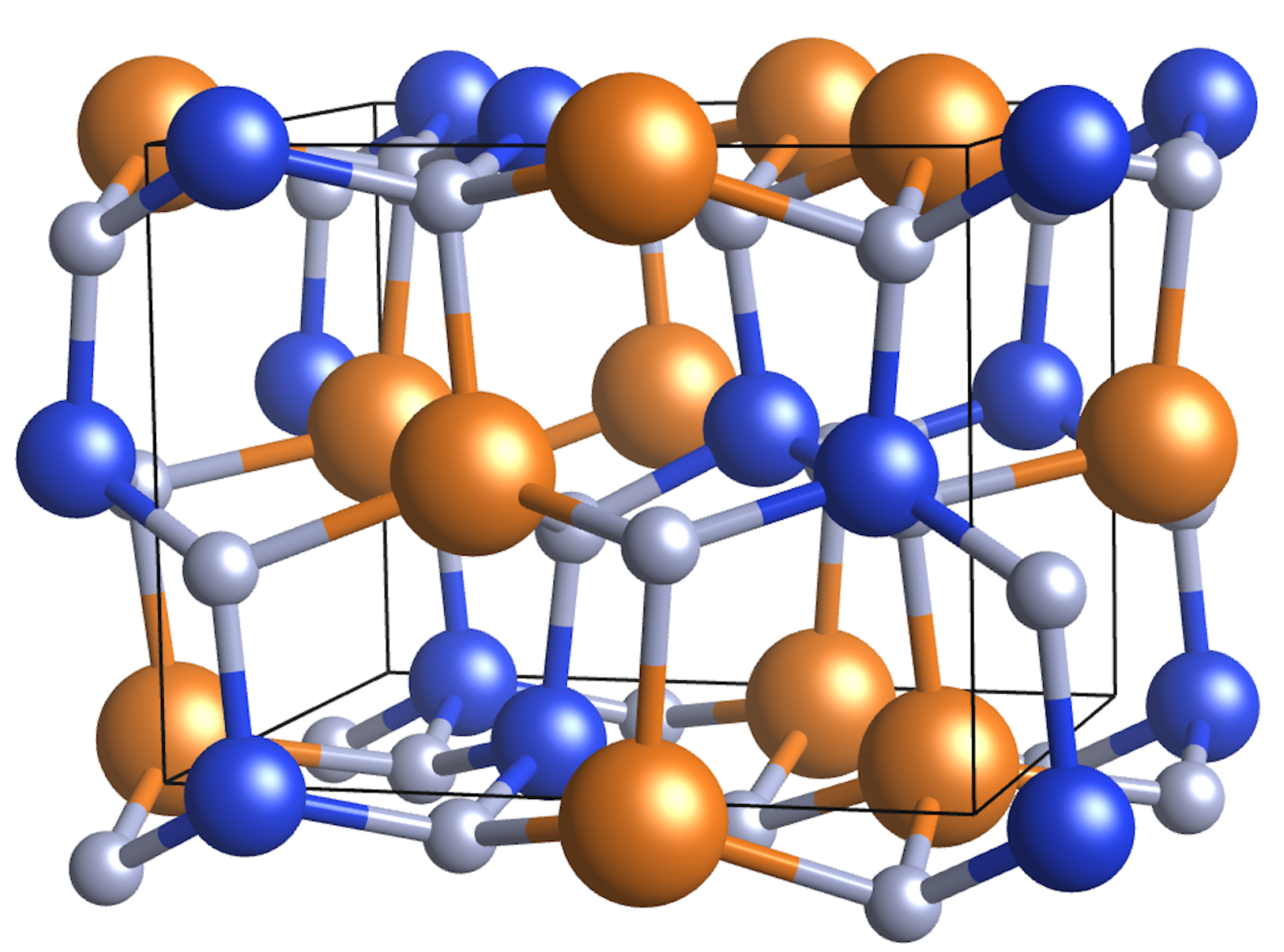}
i
\caption{\label{fig:alpha} (Color online) The crystal structure of the orthorhombic phase of MgSiN$_{2}$. Mg, Si and N are shown in bronze, blue and grey spheres respectively. The solid black lines show the boundaries of the orthorhombic unit cell.}
\end{figure}
\begin{figure}[t]
\includegraphics[width=6.0cm]{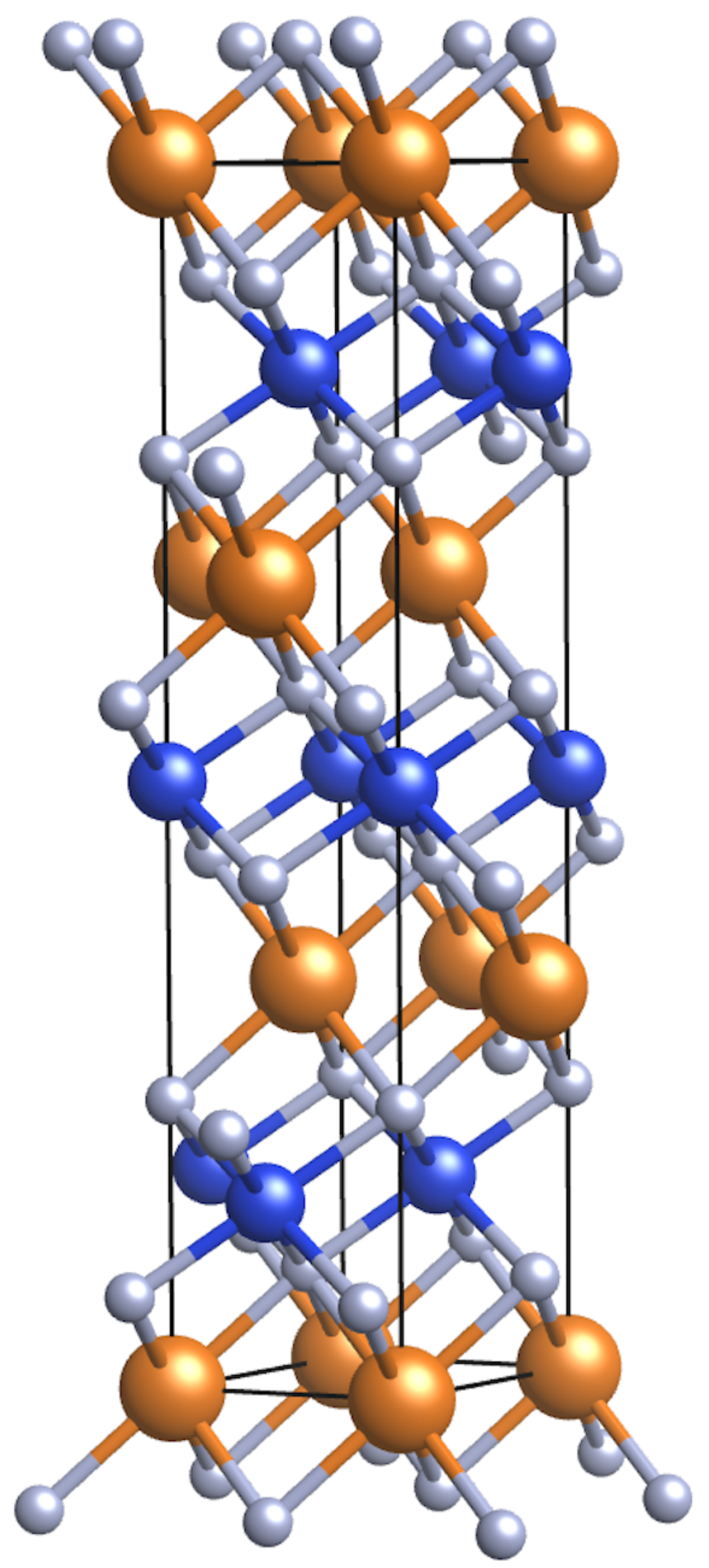}
\caption{\label{fig:beta} (Color online) The crystal structure of the high pressure rhombohedral phase of MgSiN$_{2}$ shown in its hexagonal representation. Mg, Si and N are shown in bronze, blue and grey spheres respectively. The solid black lines show the boundaries of the hexagonal unit cell.}
\end{figure}

\section{Details of the calculations}\label{sec:details}
Density functional calculations have been performed using the projector augmented wave (PAW) method\cite{Blochl} as implemented in the Vienna {\it ab initio} simulation package (VASP).\cite{KresseandFurth,KresseandJoubert} We have made use of the local density approximation (LDA) and two types of generalised gradient approximations, namely PBE\cite{PBE} and PBEsol\cite{PBEsol}. Since we are interested in an accurate determination of the electronic properties, an accurate determination of the band gap in particular, in addition to structural and energetic properties, we have also made use of the hybrid density functional approximation of Heyd, Scuseria and Ernzerhof (HSE).\cite{Heyd2003,Heyd2006} The HSE approximation has been found to significantly improve the accuracy of calculated band gaps compared to local and semi-local density functional approximations,\cite{Heyd2003,Heyd2006,Paier2005,Matsushita2011} while providing accurate energetics\cite{Paier2005} and structural properties.\cite{Paier2005,Rasander2015} The HSE approximation depends on two parameters $\alpha$ and $\omega$, where $\alpha$ determines the amount of Hartree-Fock exchange that is being used and $\omega$ determines the spatial range over which the non-local exchange is important. Here we used the usual value for $\alpha$ of 0.25, while the screening parameter $\omega$ was set to 0.2~\AA$^{-1}$.\cite{Heyd2006,Paier2005} This combination is often presented as HSE06.
\par
The plane wave energy cut-off was set to 800~eV and we have used $\Gamma$-centered k-point meshes with the smallest allowed spacing between k-points of 0.1~\AA$^{-1}$ (0.4~\AA$^{-1}$ when using the HSE approximation).  The atomic positions and simulation cell shapes were relaxed until the Hellmann-Feynman forces acting on atoms were smaller than 0.001~eV/\AA. In order to calculate the physical properties of the two phases for different pressures, we applied a series of pressures from 0~to~30~GPa and for each pressure we allowed the lattice constants and atomic coordinates to relax until the above mentioned condition on the Hellmann-Feynman forces were met. Hence, we have not used any fitting to an analytical equation of state. In the case of Mg, we have made use of PAW potentials that have the Mg 2p semicore states treated in the valence whenever the LDA, PBE and PBEsol approximations have been used. In general, the difference between treating the Mg 2p states in the valence or as cores states is small. When using the HSE approximation, the Mg 2p states have not been treated as valence states. For Si and N the standard valence treatment has been used for all exchange-correlation approximations. 
\par
\begin{table}[t]
\caption{\label{tab:structure} Ground state structural properties of LP-MgSiN$_{2}$. Wyckoff positions $(x,y,z)$ for Mg, Si and the two inequivalent N positions are shown. }
\begin{ruledtabular}
\begin{tabular}{lccc}
 & $a$ (\AA) & $b$ (\AA) & $c$ (\AA) \\
\hline
LDA & 5.240 & 6.419 & 4.957\\
PBE & 5.314 & 6.507 & 5.033\\
PBEsol & 5.280 & 6.480 & 4.999 \\
HSE & 5.264 & 6.443 & 4.978 \\
Expt.\cite{Bruls1} & 5.27078(5) & 6.46916(7) & 4.98401(5)\\
Expt.\cite{Quirk} & 5.314 & 6.466 & 4.975\\
\hline
   \multicolumn{4}{c}{LDA} \\
  & $x$ & $y$ & $z$ \\
 \hline
Mg & 0.0838 & 0.6228 & 0.9884 \\
Si &   0.0707 & 0.1255 & 0.9999\\
N(1) &   0.0491 & 0.0959 & 0.3479\\
N(2) &   0.1084 & 0.6551 & 0.4098\\
 \hline
   \multicolumn{4}{c}{PBE} \\
  & $x$ & $y$ & $z$ \\
 \hline
Mg & 0.0847 & 0.6228 & 0.9886 \\
Si & 0.0698 & 0.1254 & 0.0001\\
N(1) & 0.0480 & 0.0952 & 0.3467\\
N(2) & 0.1097 & 0.6558 & 0.4107\\
 \hline
    \multicolumn{4}{c}{PBEsol} \\
  & $x$ & $y$ & $z$ \\
 \hline
Mg & 0.0840 & 0.6227 & 0.9883 \\
Si & 0.0701 & 0.1255 & 0.0000 \\
N(1) & 0.0481 & 0.0955 & 0.3473 \\
N(2) & 0.1090 & 0.6556 & 0.4105\\
 \hline
     \multicolumn{4}{c}{HSE} \\
  & $x$ & $y$ & $z$ \\
 \hline
Mg & 0.0846 & 0.6228 &  0.9881\\
Si & 0.0701 & 0.1255 & 0.0000 \\
N(1) & 0.0485 & 0.0954 & 0.3472 \\
N(2) & 0.1093 &  0.6555 & 0.4106\\
 \hline
 \multicolumn{4}{c}{Expt.\cite{Bruls1}}\\
& $x$ & $y$ & $z$ \\
\hline
Mg & 0.08448(34) & 0.62255(30) & 0.9866(5) \\
Si & 0.0693(5) & 0.1249(4) & 0.0000\\
N(1) & 0.04855(17) & 0.09562(15) & 0.3472(4)\\
N(2) & 0.10859(18) & 0.65527(14) & 0.4102(4)\\
 \end{tabular}
\end{ruledtabular}
\end{table}

\section{Structural properties}\label{sec:structure}

LP-MgSiN$_{2}$ has an orthorhombic crystal structure, illustrated in Fig.~\ref{fig:alpha}, which can be described as a $2\times\sqrt{3}$ superlattice of the wurtzite crystal structure with the $c$-axis being common to both lattices. It has 4 formula units of MgSiN$_{2}$ per unit cell. The positions of the four types of atoms (Mg, Si, and two types of N positions) depend on the general coordinates $(x,y,z)$.  We follow the normal lattice vector convention for these materials such that $|\bar{a}_{3}|<|\bar{a}_{1}|<|\bar{a}_{2}|$. Note that in general for orthorhombic structures the convention is to have $|\bar{a}_{1}|<|\bar{a}_{2}|<|\bar{a}_{3}|$.\cite{Setyawan2010} However, our choice assures that $\bar{a}_{3}$ remains common with the $c$-axis in the wurtzite crystal structure. This choice will affect the labels of high symmetry points in the Brillouin zone when comparing results obtained using the two different conventions.
\par
The calculated structural properties of LP-MgSiN$_{2}$ are shown in Table~\ref{tab:structure}. As can be seen, the calculated lattice constants and atomic positions are in very good agreement with available experimental structures, with deviations in the lattice constants of about 1\% or better, as expected for these exchange-correlation approximations.\cite{Rasander2015}
\par
The high pressure phase has a rhombohedral structure, illustrated in its hexagonal representation in Fig.~\ref{fig:beta}, where Mg and Si occupies the 3a and 3b positions with coordinates $(0,0,0)$ and $(0,0,1/2)$, respectively. N occupies the 6c positions with coordinate $(0,0,z)$. The calculated structural parameters are shown in Table~\ref{tab:structure-beta}. Once again, the agreement between theory and experiment is very good with an error in the lattice constants smaller than 1\%.

\begin{table}[t]
\caption{\label{tab:structure-beta} Ground state structural properties of HP-MgSiN$_{2}$. Mg and Si atoms are found at the $(0,0,0)$ and $(0,0,1/2)$ positions respectively. The $z$ parameter for the N positions $(0,0,z)$ is also shown.}
\begin{ruledtabular}
\begin{tabular}{lcccc}
 & $a$ (\AA) & $c$ (\AA) & $c/a$ & $z$\\
\hline
LDA & 2.810 & 14.461 & 5.147 &  0.2369\\
PBE & 2.852 & 14.699 & 5.154 & 0.2366\\
PBEsol & 2.833 & 14.604 & 5.155 & 0.2367\\
HSE & 2.817 & 14.510 & 5.151 & 0.2367\\
Expt.\cite{Andrade2011} & 2.8383(3) & 14.558(2) & 5.129(1) & 0.2385(3) \\
 \end{tabular}
\end{ruledtabular}
\end{table}
\begin{figure*}[t]
\includegraphics[width=16cm]{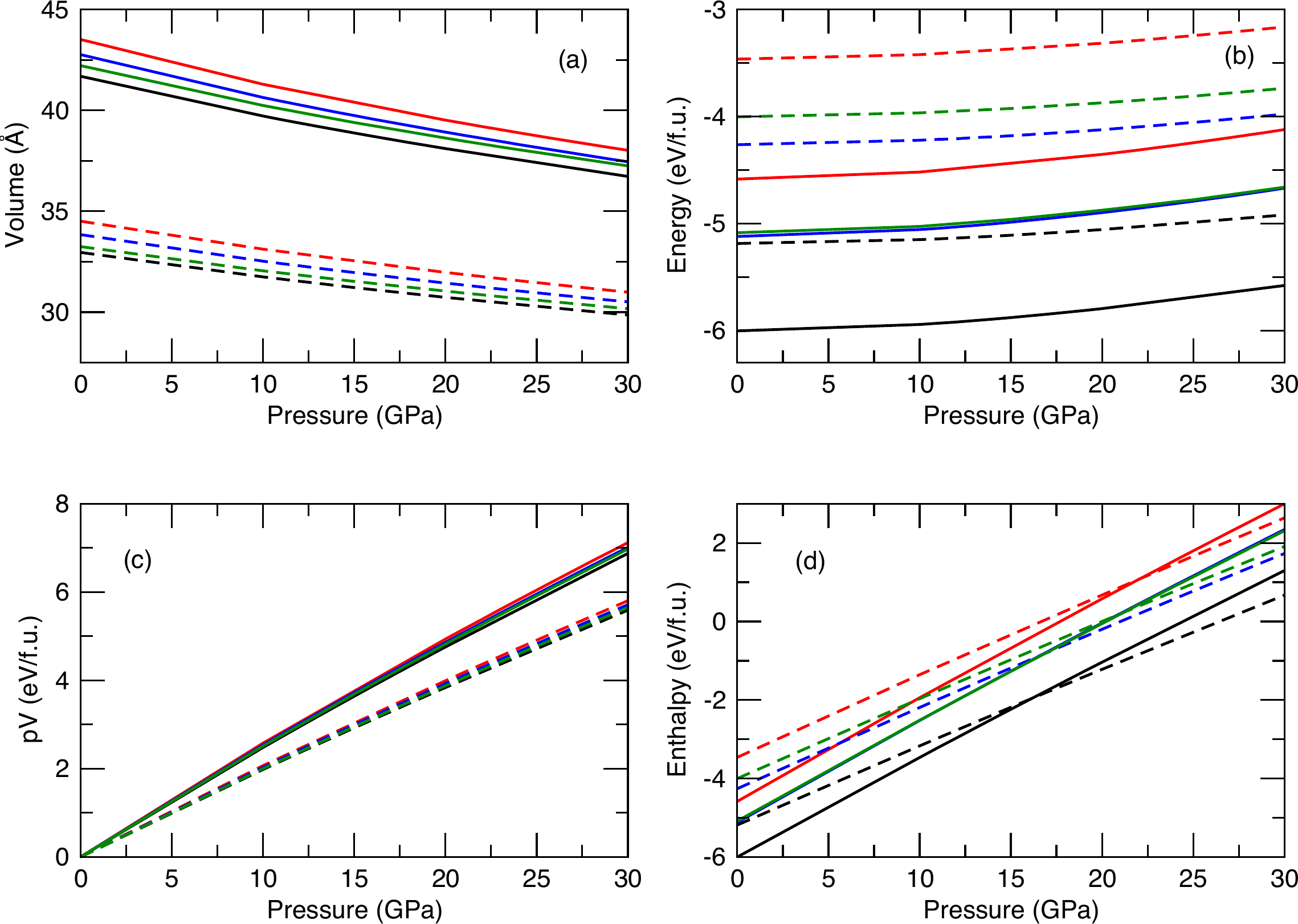}
\caption{\label{fig:properties} (Color online) Calculated volume (a), total energies (b), pressure-volume terms (c) and enthalpies (d) for LP- (solid lines) and HP-phases (dashed lines) of MgSiN$_{2}$. Results obtain using the LDA, PBE, PBEsol and HSE approximations are shown using black, red, blue and green colour, respectively.}
\end{figure*}

\section{Phase stability}\label{sec:stability}
\par

\begin{figure}[t]
\includegraphics[width=8cm]{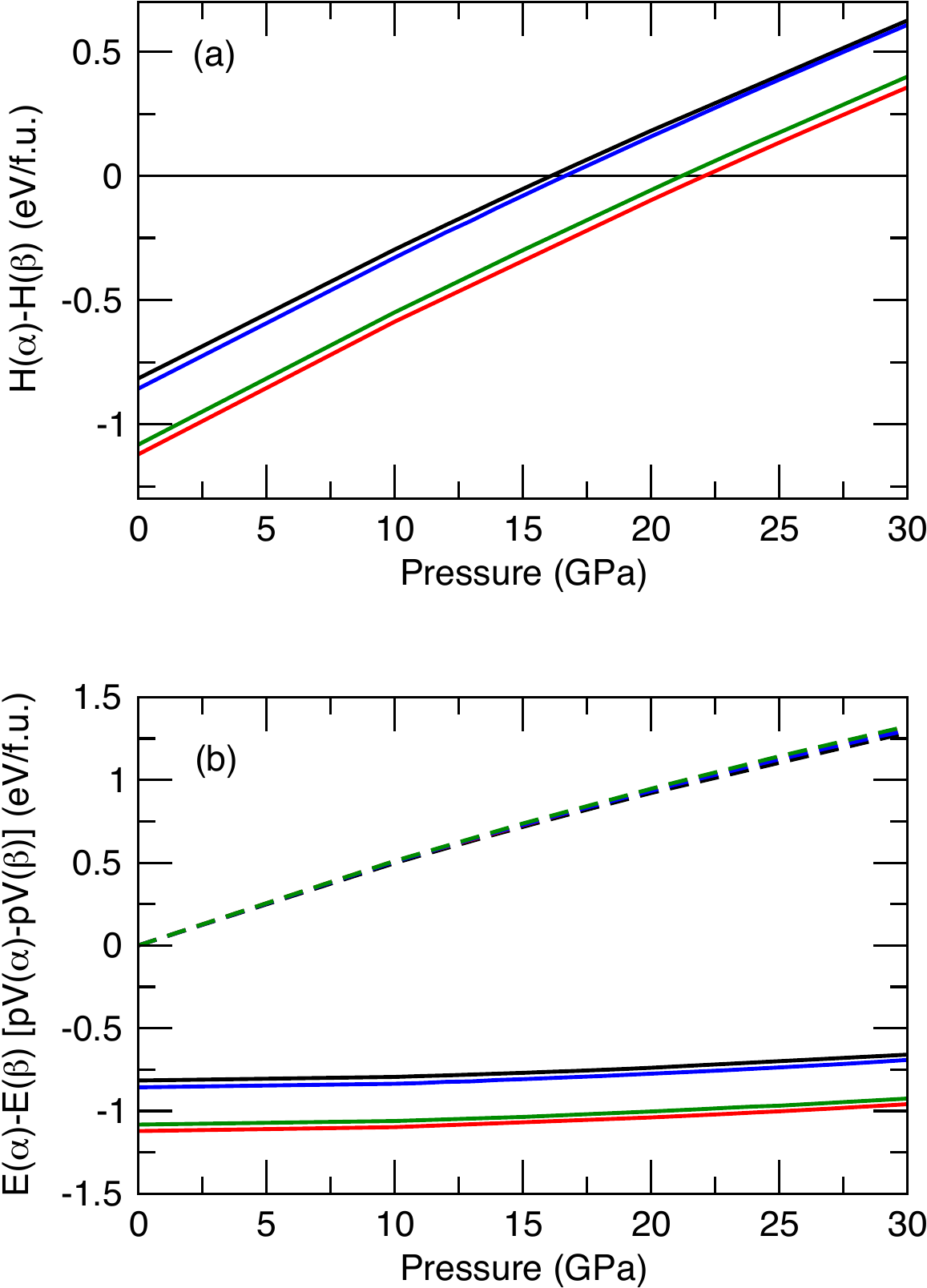}
\caption{\label{fig:H} (Color online) Calculated enthalpy differences between the LP- and HP-phases of MgSiN$_{2}$ (a) and calculated differences (b) in the total energy, $E$ (solid lines), and the pressure-volume term, $pV$ (dashed lines), in Eqn.~(\ref{eq:H}). Results from the LDA, PBE, PBEsol and HSE approximations are shown in black, red, blue and green, respectively. The phase transition between LP- and HP-phases occur when the calculated enthalpy differences cross the zero line, marked by the vertical solid line. The transition pressure is evaluated to be 16~GPa (LDA), 22~GPa (PBE), 16.5~GPa (PBEsol) and 21~GPa (HSE). }
\end{figure}

We have evaluated the relative stability of MgSiN$_{2}$ under pressure by calculating the enthalpy, $H$, as
\begin{equation}\label{eq:H}
H=E+pV,
\end{equation}
where $E$ is the total energy and $V$ is the volume of the system, for each phase at pressures ($p$) between 0 and 30 GPa for both phases of MgSiN$_{2}$. In Fig.~\ref{fig:properties} we show the calculated volumes, total energies, pressure-volume ($pV$) terms, and enthalpies for the two phases. All energies are shown with respect to each elements standard reference state at $p=0$ and have been evaluated as
\begin{equation}
 E({\rm MgSiN}_{2}) - \big(E({\rm Mg}) + E({\rm Si}) + E({\rm N}_{2})\big),
\end{equation}
where $E({\rm Mg})$, $E({\rm Si})$ and $E({\rm N}_{2})$ are the energies of hexagonal close-packed Mg, diamond structured Si and the N$_{2}$ molecule, respectively, and $E({\rm MgSiN}_{2})$ is the energy of of one formula unit of MgSiN$_{2}$ . In addition, we provide in Fig.~\ref{fig:H} the calculated differences in the enthalpies, total energies, and pressure-volume terms between the two MgSiN$_{2}$ phases. The enthalpy of both phases, shown in Fig.~\ref{fig:properties}, shows an almost linear increase with pressure. Initially, the enthalpy is lower for the LP-phase than for the HP-phase. The enthalpy difference at $p=0$ is 0.82~eV (LDA), 1.12~eV (PBE), 0.86~eV (PBEsol) and 1.08~eV (HSE), which is in agreement with previous studies.\cite{Romer2009,Arab2015} When the pressure increases, the enthalpy of the HP-phase increase slower compared to the LP-phase and a phase transition occurs at 16~GPa (LDA), 22~GPa (PBE), 16.5~GPa (PBEsol) and 21~GPa (HSE), when the rhombohedral phase has the lower enthalpy, which is clearly shown in Figs.~\ref{fig:properties}~and~\ref{fig:H}. R{\"o}mer {\it et al.} found the transition to occur at 25~GPa using PW91,\cite{Romer2009} Fang {\it et al.} reported the transition to occur at 16.5~GPa using the LDA,\cite{Fang2} while Arab {\it et al.}\cite{Arab2015} recently found the transition pressure to be 25~GPa, 17.45~GPa and 19.05~GPa using the PW91, PBEsol and LDA approximations, respectively. The experimentally evaluated transition pressure is 27~GPa.\cite{Andrade2011} Our calculations are therefore in good agreement with previous theoretical studies as well as with available experiments. Note that none of the mentioned calculations (including our results) include any temperature effects and that the experimental transition pressure was obtained using a high temperature (exceeding 2000~K\cite{Andrade2011}). A perfect agreement with experiment would therefore be coincidental. 

\par
We note that the evaluated phase transition pressures follow the calculated enthalpy differences between the LP- and HP-phases at $p=0$, since a larger evaluated enthalpy difference at $p=0$ provides a higher transition pressure and vice versa. This behaviour can be explained by analysing the difference in enthalpy, total energy and pressure-volume characteristics between the LP- and HP-phases, which is shown in Fig.~\ref{fig:H}. As can be seen in Fig.~\ref{fig:H}, the difference in total energy varies slightly with pressure and always shows a lower energy for the LP-phase irrespective of the exhange-correlation approximation. The difference in the pressure-volume term of Eqn.~(\ref{eq:H}) show a much larger increase with pressure, but hardly any difference between using the different exchange-correlation approximations. The reason for the phase transition is therefore to be found in the contribution to the enthalpy due to the pressure-volume characteristics of the two phases, which is clearly illustrated in Fig.~\ref{fig:H}. The HP-phase has a much smaller volume than the LP-phase, which gives a smaller $pV$-contribution to the enthalpy of the HP-phase and lower increase of the enthalpy with pressure compared to the LP-phase.
\par
\begin{figure}[t]
\includegraphics[width=9cm]{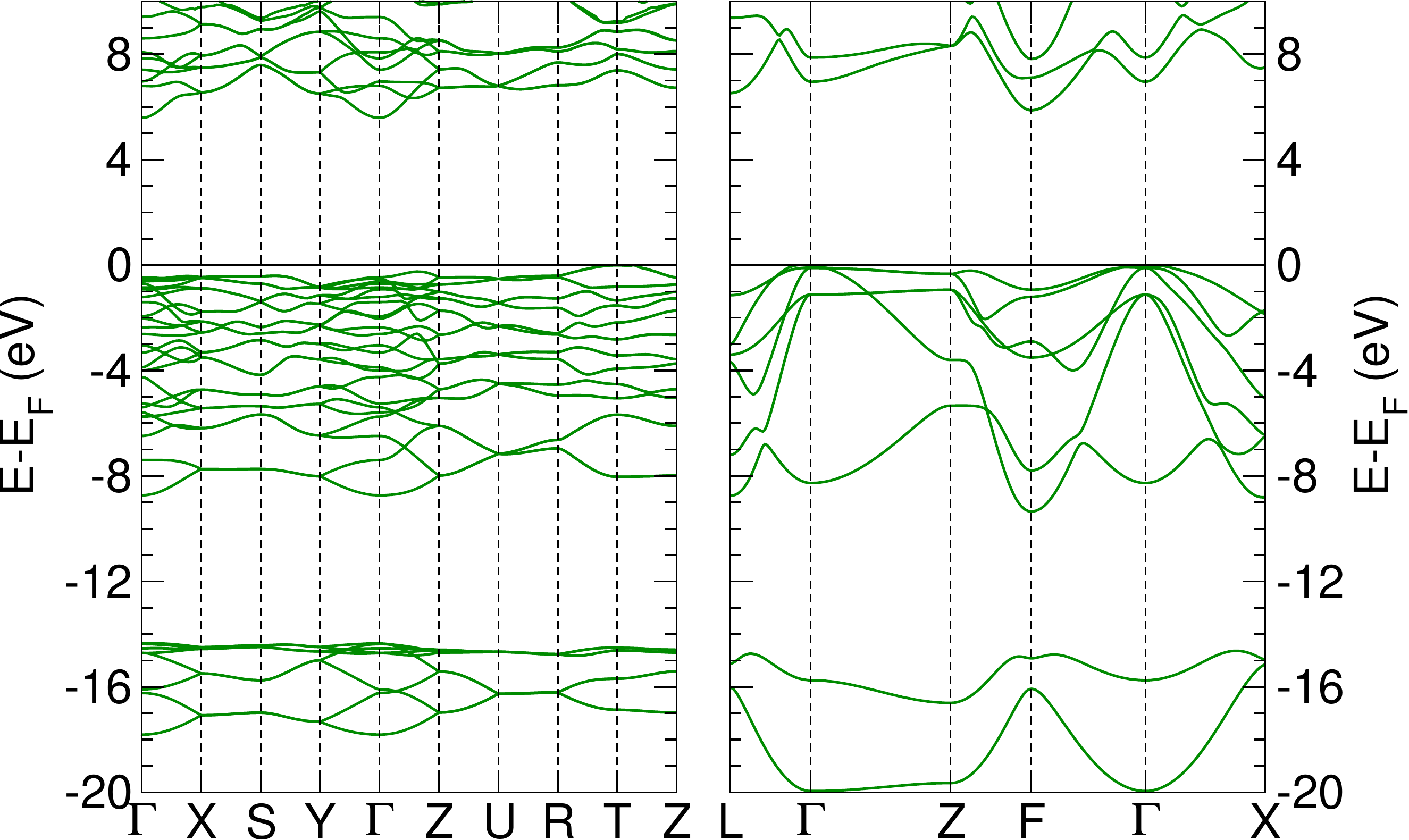}
\caption{\label{fig:bands} (Color online) Calculated energy bands along high symmetry direction in the Brillouin zone for LP- (left) and HP-MgSiN$_{2}$ (right) obtained using the HSE approximation. All energies are in relation to the Fermi level, $E_{F}$.}
\end{figure}

\section{Electronic structure}\label{sec:electronic}
\par
In Fig.~\ref{fig:bands} we show the calculated ground state band structures obtained using the HSE approximation. LP-MgSiN$_{2}$ has a large indirect band gap of 5.58~eV in agreement with experimental observations.\cite{deBoer2015} The valence band maximum (VBM) is positioned at the T-point and the conduction band minimum (CBM) is found at the $\Gamma$-point. The VBM is doubly degenerate and the CBM is non-degenerate. The CBM at the $\Gamma$-point extends more than 1~eV lower than the other conduction band states at higher energies. We note that the main features of the band structure are very similar when comparing the LDA, PBE, PBEsol and HSE approximations, apart from the size of the band gap which is found to be 4.31~eV, 3.95~eV and 4.02~eV for the LDA, PBE and PBEsol approximations, respectively, and therefore significantly smaller for the local and semilocal approximations compared to the HSE approximation.
\par
\begin{figure}[b]
\includegraphics[width=7cm]{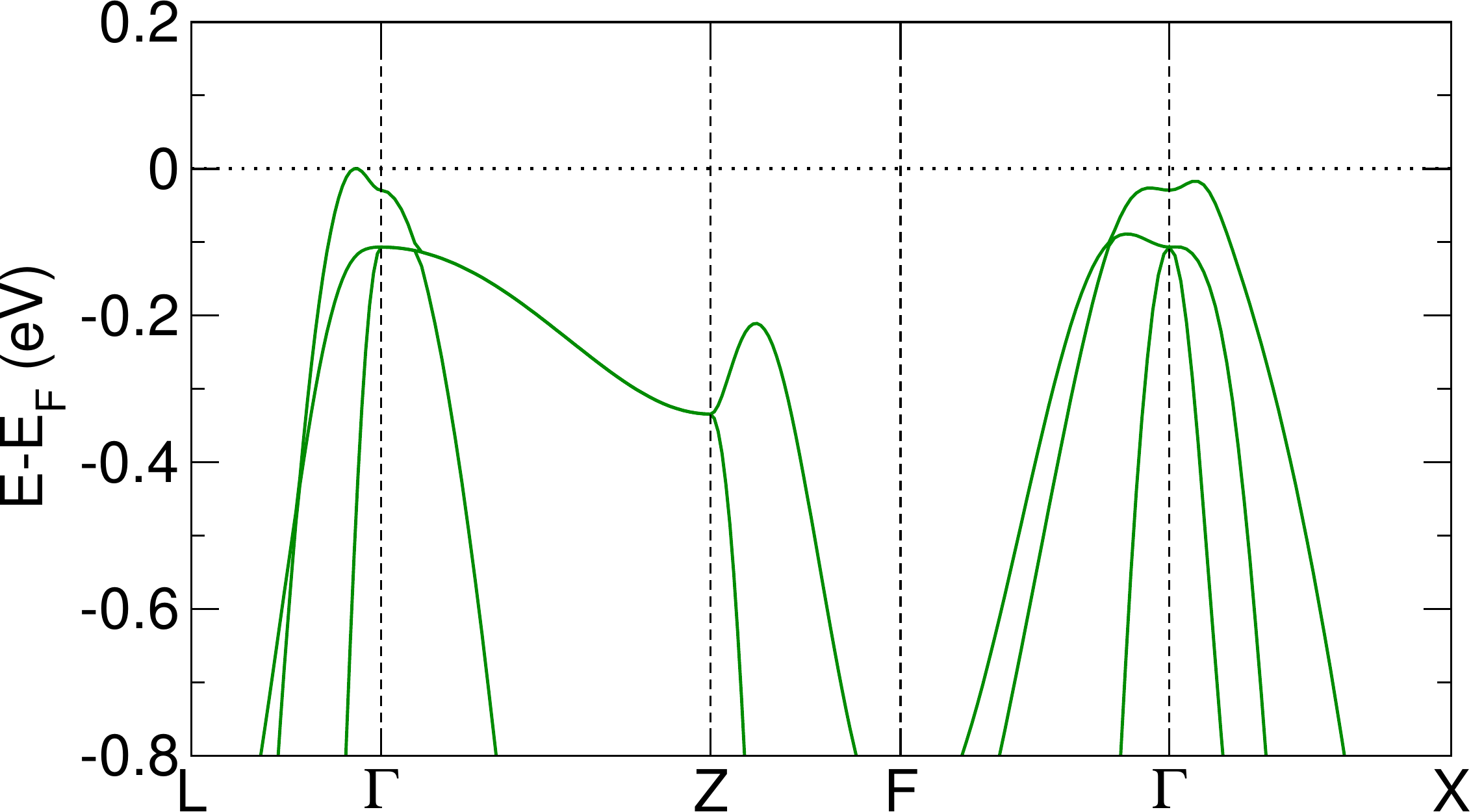}
\caption{\label{fig:betabands} (Color online) Calculated energy bands along high symmetry directions in the Brillouin zone of the HP-phase of MgSiN$_{2}$ close to the Fermi level, $E_{F}$, as obtained using the HSE approximation.}
\end{figure}

The band gap of the HP-phase within the HSE approximation is 5.87~eV while the LDA, PBE and PBEsol band gaps are 4.37~eV, 4.16~eV and 4.25~eV, respectively. The VBM is found slightly off the $\Gamma$-point towards the L-point, with the Brillouin zone coordinates $(\delta,0,0)$, with $\delta\approx0.07$. Furthermore, we note that the highest valence band increases its energy when moving from the $\Gamma$-point towards the L-, F- and X-points, where local maximas are found slightly off the $\Gamma$-point in all directions, while the energy of the band decrease when moving from the $\Gamma$-point towards the Z-point. We note that the energies of the highest valence band at the two local maximas along $\Gamma$-L and $\Gamma$-X are very similar. This is shown more clearly in Fig.~\ref{fig:betabands} where the band structure close to the Fermi level of HP-MgSiN$_{2}$ is shown. Furthermore, we also find that the band gap of the HP-phase is indirect with the CBM at the F-point.

\begin{figure}[t]
\includegraphics[width=9cm]{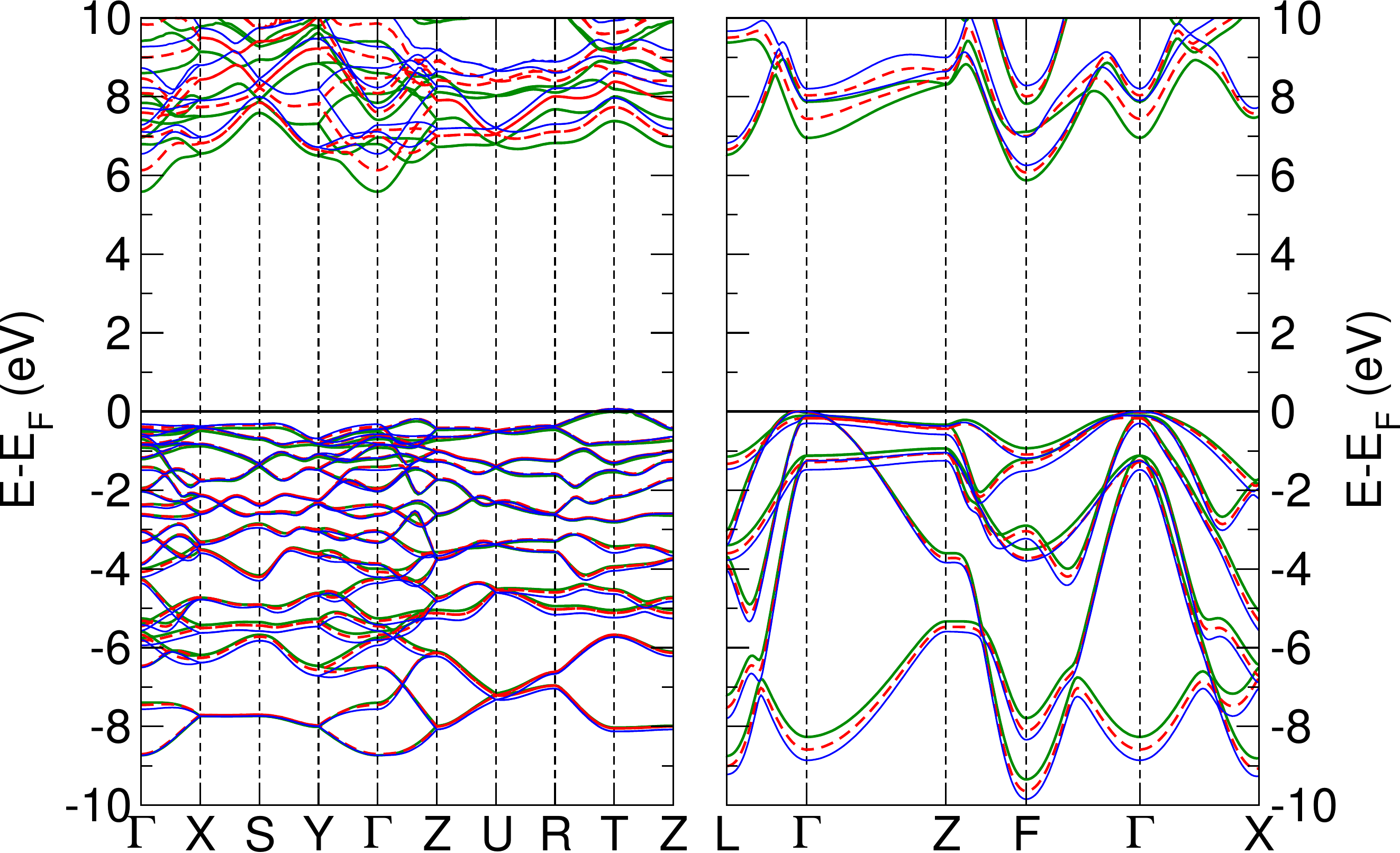}
\caption{\label{fig:bands-pressure} (Color online) Calculated energy bands along high symmetry direction in the Brillouin zone for LP- (left) and HP-MgSiN$_{2}$ (right) for $p=0$~(solid green line), 10 GPa (dashed red line) and 20 GPa (thin solid blue line) obtained using the HSE approximation. All energies are in relation to the Fermi level, $E_{F}$.}
\end{figure}
\par
We note that HP-MgSiN$_{2}$ has a larger band gap in the ground state compared to the LP-phase; the difference is 0.29~eV as obtained by the HSE approximation. In Fig.~\ref{fig:bands-pressure}, we show the calculated energy bands as a function of pressure. When the pressure increases the valence and conduction bands become more separated and consequently the band gaps increase with pressure. Furthermore, the band gap in the low pressure phase increases more with pressure than the high pressure phase. This change is largely due to the CBM at the $\Gamma$-point in the LP-phase rising compared to the other conduction bands and becoming more or less degenerate with the other conduction bands at higher pressures. In the HP-phase the conduction bands move more collectively away from the valence bands as the pressure increases. As a consequence, the band gap in the LP-phase becomes larger than in the HP-phase. This change occur well below the phase transition pressure of the system, since it is found already for $p=10$~GPa, as shown in Fig.~\ref{fig:bands-pressure}.
\begin{figure*}[t]
\includegraphics[width=6cm]{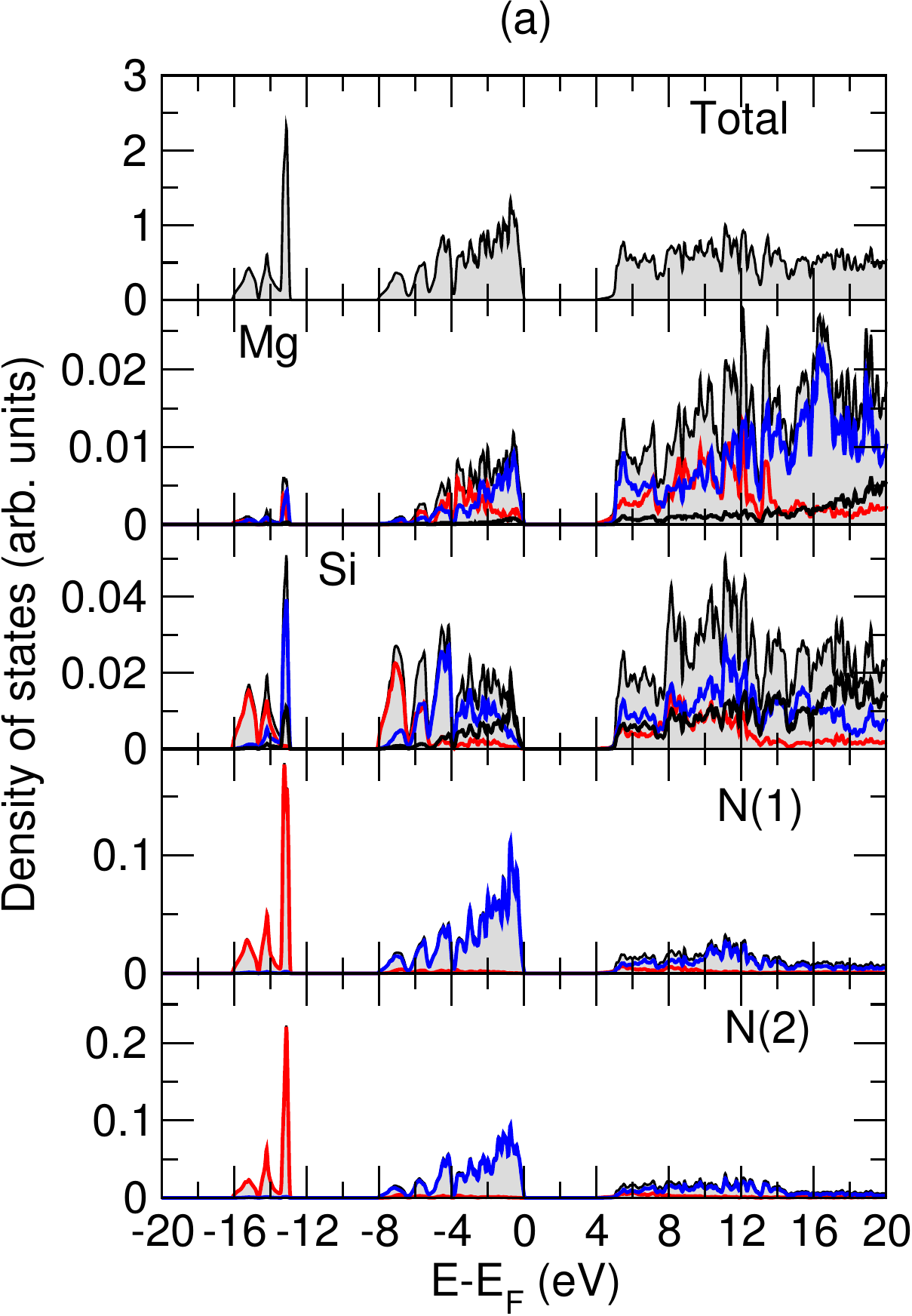}
\includegraphics[width=6cm]{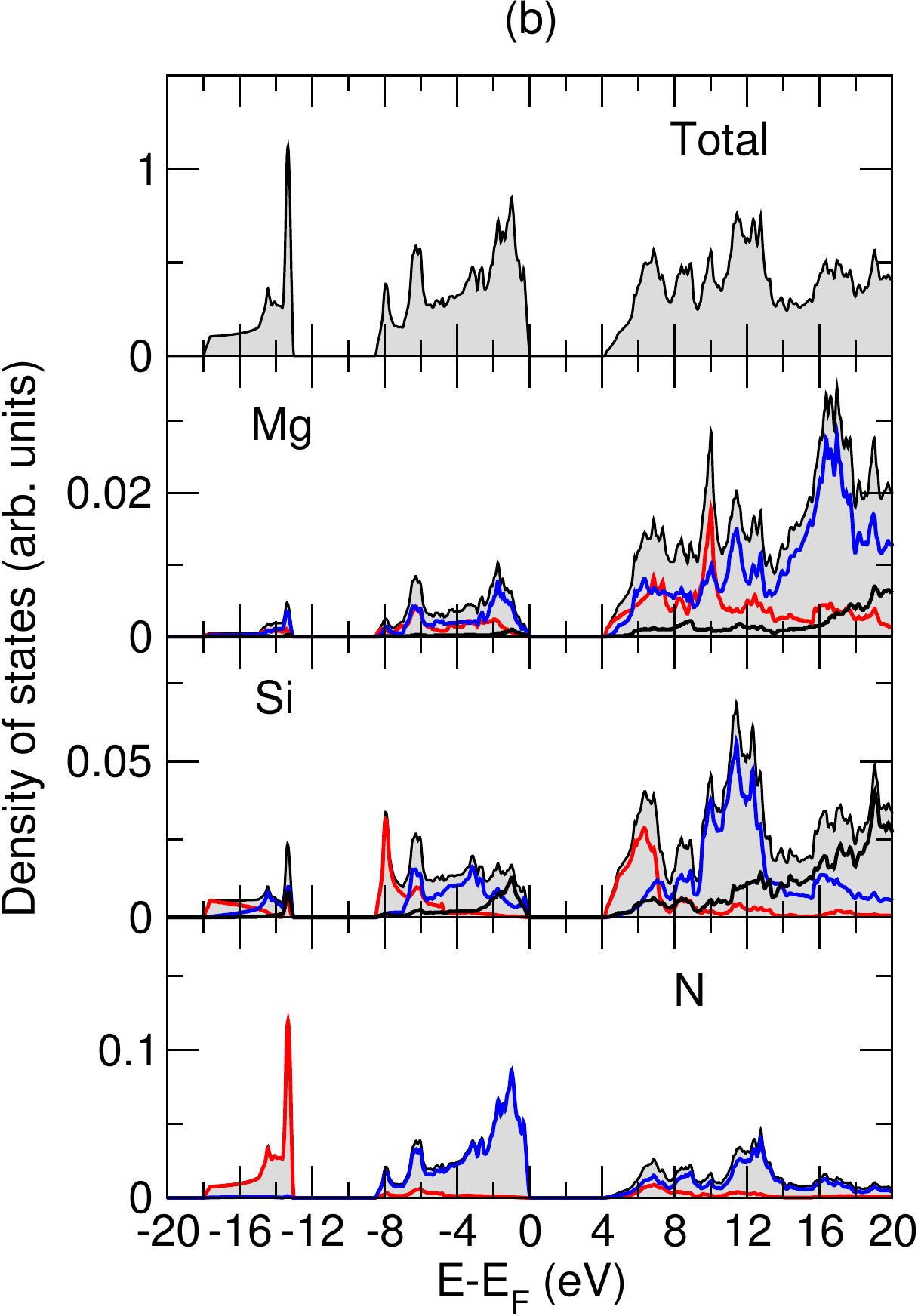}
\caption{\label{fig:dos} (Color online) Calculated density of states (DOS) of LP- (a) and HP-phases (b) of MgSiN$_{2}$. The total DOS and the DOS projected on to Mg, Si and N atoms are shown (black lines with grey shading) as well as the projection onto the s- (red), p- (blue) and d-states (black) of each atomic species. The DOS shown are obtained using the PBE approximation. All energies are in relation to the Fermi level, $E_{F}$.}
\end{figure*}
\par
In Fig.~\ref{fig:dos}, we show the calculated density of states (DOS) of LP- and HP-MgSiN$_{2}$ as obtained within the PBE approximation. Beginning with the LP-phase, we find that states close to the Fermi-level are dominated by N. In the region between -8~eV to the Fermi-level, the N states are completely dominated by N p-states. These states are to a large part mixed with Mg and Si p-states. However, there are contributions coming from N p-states mixed with Mg and Si s-states and also some Mg and Si states with d-character are mixed with the N p-states in the upper part of the valence band. In the lower part of the valence band between -16~eV and -13~eV, the N states have s-character. These states are mixed with Mg and Si s- and p-states. Overall, the features in the DOS of the LP-phase is similar to the DOS of wurtzite AlN.\cite{Quirk}
\par
For the HP-phase, we find a rather similar DOS as for the LP-phase, even though the atoms are now six-fold coordinated in contrast to the four-fold coordinated LP-phase. Compared to the DOS of the LP-phase the valence bands in the HP-phase are wider, especially the lower valence band region which contain N s-states mixed with Mg and Si s- and p-states are found between -18~eV and~-13~eV. In the region between -8~eV and the Fermi-level we find, as in the LP-phase, N p-states mixed with Mg and Si s-, p- and d-states.

\section{Conclusions}\label{sec:conclusions}
In conclusion, we have calculated the phase stability under pressure of two MgSiN$_{2}$ phases and found that there is a phase transition at a pressure of about 20~GPa, from the ground state orthorhombic phase (space group Pna2$_{1}$) to the high pressure phase of rhombohedral geometry (space group R$\bar{3}$m) in agreement with previous theory and experiment. We note that the difference obtained using various approximations for the exchange-correlation energy for the phase transition pressure depends strongly on the total energy difference at $p=0$. If the difference in energy for a specific XC approximation between LP- and HP-phases at $p=0$ is large this moves the phase transition pressure towards higher pressures and vice versa. 
\par
The electronic structures of these two phases have been calculated using hybrid density functional theory. Both phases are wide band gap semiconductors with indirect band gaps. For the ground states the HP-phase has a larger gap of 5.87~eV while the LP-phase has a gap of 5.58~eV. However, as the pressure increases the order is shifted and at the transition pressure the LP-phase has the larger band gap. This change is due to the CBM in the LP-phase becoming aligned with other higher energy conduction bands, while the conduction bands in the HP-phase shifts away from the valence bands more collectively.

\section{Acknowledgements}
We acknowledge support from the Leverhulme Trust via M. A. Moram's Research Leadership Award (RL-007-2012). M. A. Moram acknowledges further support from the Royal Society through a University Research Fellowship. This work used the Imperial College high performance computing facilities and, via our membership of the UK's HEC Materials Chemistry Consortium funded by EPSRC (EP/L000202), the ARCHER UK National Supercomputing Service (http://www.archer.ac.uk).

\bibliography{mgsin2}

\begin{thebibliography}{23}%
\makeatletter
\providecommand \@ifxundefined [1]{%
 \@ifx{#1\undefined}
}%
\providecommand \@ifnum [1]{%
 \ifnum #1\expandafter \@firstoftwo
 \else \expandafter \@secondoftwo
 \fi
}%
\providecommand \@ifx [1]{%
 \ifx #1\expandafter \@firstoftwo
 \else \expandafter \@secondoftwo
 \fi
}%
\providecommand \natexlab [1]{#1}%
\providecommand \enquote  [1]{``#1''}%
\providecommand \bibnamefont  [1]{#1}%
\providecommand \bibfnamefont [1]{#1}%
\providecommand \citenamefont [1]{#1}%
\providecommand \href@noop [0]{\@secondoftwo}%
\providecommand \href [0]{\begingroup \@sanitize@url \@href}%
\providecommand \@href[1]{\@@startlink{#1}\@@href}%
\providecommand \@@href[1]{\endgroup#1\@@endlink}%
\providecommand \@sanitize@url [0]{\catcode `\\12\catcode `\$12\catcode
  `\&12\catcode `\#12\catcode `\^12\catcode `\_12\catcode `\%12\relax}%
\providecommand \@@startlink[1]{}%
\providecommand \@@endlink[0]{}%
\providecommand \url  [0]{\begingroup\@sanitize@url \@url }%
\providecommand \@url [1]{\endgroup\@href {#1}{\urlprefix }}%
\providecommand \urlprefix  [0]{URL }%
\providecommand \Eprint [0]{\href }%
\providecommand \doibase [0]{http://dx.doi.org/}%
\providecommand \selectlanguage [0]{\@gobble}%
\providecommand \bibinfo  [0]{\@secondoftwo}%
\providecommand \bibfield  [0]{\@secondoftwo}%
\providecommand \translation [1]{[#1]}%
\providecommand \BibitemOpen [0]{}%
\providecommand \bibitemStop [0]{}%
\providecommand \bibitemNoStop [0]{.\EOS\space}%
\providecommand \EOS [0]{\spacefactor3000\relax}%
\providecommand \BibitemShut  [1]{\csname bibitem#1\endcsname}%
\let\auto@bib@innerbib\@empty
\bibitem [{\citenamefont {Punya}\ \emph {et~al.}()\citenamefont {Punya},
  \citenamefont {Paudel},\ and\ \citenamefont {Lambrecht}}]{Punya2011}%
  \BibitemOpen
  \bibfield  {author} {\bibinfo {author} {\bibfnamefont {A.}~\bibnamefont
  {Punya}}, \bibinfo {author} {\bibfnamefont {T.~R.}\ \bibnamefont {Paudel}}, \
  and\ \bibinfo {author} {\bibfnamefont {W.~R.~L.}\ \bibnamefont {Lambrecht}},\
  }\href@noop {} {\bibfield  {journal} {\bibinfo  {journal} {Phys. Status.
  Solidi C}\ }\textbf {\bibinfo {volume} {8}},\ \bibinfo {pages}
  {2492}}\BibitemShut {NoStop}%
\bibitem [{\citenamefont {David}\ \emph {et~al.}(1970)\citenamefont {David},
  \citenamefont {Laurent},\ and\ \citenamefont {Lang}}]{David1970}%
  \BibitemOpen
  \bibfield  {author} {\bibinfo {author} {\bibfnamefont {J.}~\bibnamefont
  {David}}, \bibinfo {author} {\bibfnamefont {Y.}~\bibnamefont {Laurent}}, \
  and\ \bibinfo {author} {\bibfnamefont {J.}~\bibnamefont {Lang}},\ }\href@noop
  {} {\bibfield  {journal} {\bibinfo  {journal} {Bull. Soci{\'e}t{\'e} Fr.
  Min{\'e}ralogie Cristallogr.}\ }\textbf {\bibinfo {volume} {93}},\ \bibinfo
  {pages} {153} (\bibinfo {year} {1970})}\BibitemShut {NoStop}%
\bibitem [{\citenamefont {Bruls}\ \emph {et~al.}(2000)\citenamefont {Bruls},
  \citenamefont {Hintzen}, \citenamefont {Metselaar},\ and\ \citenamefont
  {Loong}}]{Bruls1}%
  \BibitemOpen
  \bibfield  {author} {\bibinfo {author} {\bibfnamefont {R.~J.}\ \bibnamefont
  {Bruls}}, \bibinfo {author} {\bibfnamefont {H.~T.}\ \bibnamefont {Hintzen}},
  \bibinfo {author} {\bibfnamefont {R.}~\bibnamefont {Metselaar}}, \ and\
  \bibinfo {author} {\bibfnamefont {C.-K.}\ \bibnamefont {Loong}},\ }\href@noop
  {} {\bibfield  {journal} {\bibinfo  {journal} {J. Phys. Chem. Solids}\
  }\textbf {\bibinfo {volume} {61}},\ \bibinfo {pages} {1285} (\bibinfo {year}
  {2000})}\BibitemShut {NoStop}%
\bibitem [{\citenamefont {Quirk}\ \emph {et~al.}(2014)\citenamefont {Quirk},
  \citenamefont {R{\aa}sander}, \citenamefont {McGilvery}, \citenamefont
  {Palgrave},\ and\ \citenamefont {Moram}}]{Quirk}%
  \BibitemOpen
  \bibfield  {author} {\bibinfo {author} {\bibfnamefont {J.~B.}\ \bibnamefont
  {Quirk}}, \bibinfo {author} {\bibfnamefont {M.}~\bibnamefont {R{\aa}sander}},
  \bibinfo {author} {\bibfnamefont {C.~M.}\ \bibnamefont {McGilvery}}, \bibinfo
  {author} {\bibfnamefont {R.}~\bibnamefont {Palgrave}}, \ and\ \bibinfo
  {author} {\bibfnamefont {M.~A.}\ \bibnamefont {Moram}},\ }\href@noop {}
  {\bibfield  {journal} {\bibinfo  {journal} {Appl. Phys. Lett.}\ }\textbf
  {\bibinfo {volume} {105}},\ \bibinfo {pages} {112108} (\bibinfo {year}
  {2014})}\BibitemShut {NoStop}%
\bibitem [{\citenamefont {de~Boer}\ \emph {et~al.}(2015)\citenamefont
  {de~Boer}, \citenamefont {Boyko}, \citenamefont {Braun}, \citenamefont
  {Schnick},\ and\ \citenamefont {Moews}}]{deBoer2015}%
  \BibitemOpen
  \bibfield  {author} {\bibinfo {author} {\bibfnamefont {T.}~\bibnamefont
  {de~Boer}}, \bibinfo {author} {\bibfnamefont {T.~D.}\ \bibnamefont {Boyko}},
  \bibinfo {author} {\bibfnamefont {C.}~\bibnamefont {Braun}}, \bibinfo
  {author} {\bibfnamefont {W.}~\bibnamefont {Schnick}}, \ and\ \bibinfo
  {author} {\bibfnamefont {A.}~\bibnamefont {Moews}},\ }\href@noop {}
  {\bibfield  {journal} {\bibinfo  {journal} {Phys. Status Solidi RRL}\
  }\textbf {\bibinfo {volume} {9}},\ \bibinfo {pages} {250} (\bibinfo {year}
  {2015})}\BibitemShut {NoStop}%
\bibitem [{\citenamefont {Gaido}\ \emph {et~al.}(1974)\citenamefont {Gaido},
  \citenamefont {Dubrovskii},\ and\ \citenamefont {Zykov}}]{Gaido1974}%
  \BibitemOpen
  \bibfield  {author} {\bibinfo {author} {\bibfnamefont {G.~K.}\ \bibnamefont
  {Gaido}}, \bibinfo {author} {\bibfnamefont {G.~P.}\ \bibnamefont
  {Dubrovskii}}, \ and\ \bibinfo {author} {\bibfnamefont {A.~M.}\ \bibnamefont
  {Zykov}},\ }\href@noop {} {\bibfield  {journal} {\bibinfo  {journal} {Izv.
  Akad. Nauk. SSSR Neorg. Mater}\ }\textbf {\bibinfo {volume} {10}},\ \bibinfo
  {pages} {564} (\bibinfo {year} {1974})}\BibitemShut {NoStop}%
\bibitem [{\citenamefont {Fang}\ \emph {et~al.}(2004)\citenamefont {Fang},
  \citenamefont {Hintzen},\ and\ \citenamefont {de~With}}]{Fang2}%
  \BibitemOpen
  \bibfield  {author} {\bibinfo {author} {\bibfnamefont {C.~M.}\ \bibnamefont
  {Fang}}, \bibinfo {author} {\bibfnamefont {H.~T.}\ \bibnamefont {Hintzen}}, \
  and\ \bibinfo {author} {\bibfnamefont {G.}~\bibnamefont {de~With}},\
  }\href@noop {} {\bibfield  {journal} {\bibinfo  {journal} {Appl. Phys. A}\
  }\textbf {\bibinfo {volume} {78}},\ \bibinfo {pages} {717} (\bibinfo {year}
  {2004})}\BibitemShut {NoStop}%
\bibitem [{\citenamefont {R{\"o}mer}\ \emph {et~al.}(2009)\citenamefont
  {R{\"o}mer}, \citenamefont {Kroll},\ and\ \citenamefont
  {Schnick}}]{Romer2009}%
  \BibitemOpen
  \bibfield  {author} {\bibinfo {author} {\bibfnamefont {S.~R.}\ \bibnamefont
  {R{\"o}mer}}, \bibinfo {author} {\bibfnamefont {P.}~\bibnamefont {Kroll}}, \
  and\ \bibinfo {author} {\bibfnamefont {W.}~\bibnamefont {Schnick}},\
  }\href@noop {} {\bibfield  {journal} {\bibinfo  {journal} {J. Phys.: Condens.
  Matter}\ }\textbf {\bibinfo {volume} {21}},\ \bibinfo {pages} {275407}
  (\bibinfo {year} {2009})}\BibitemShut {NoStop}%
\bibitem [{\citenamefont {Andrade}\ \emph {et~al.}(2011)\citenamefont
  {Andrade}, \citenamefont {Dzivenko}, \citenamefont {Miehe}, \citenamefont
  {Boehler}, \citenamefont {Hintzen},\ and\ \citenamefont
  {Riedel}}]{Andrade2011}%
  \BibitemOpen
  \bibfield  {author} {\bibinfo {author} {\bibfnamefont {M.}~\bibnamefont
  {Andrade}}, \bibinfo {author} {\bibfnamefont {D.}~\bibnamefont {Dzivenko}},
  \bibinfo {author} {\bibfnamefont {G.}~\bibnamefont {Miehe}}, \bibinfo
  {author} {\bibfnamefont {R.}~\bibnamefont {Boehler}}, \bibinfo {author}
  {\bibfnamefont {H.~T.}\ \bibnamefont {Hintzen}}, \ and\ \bibinfo {author}
  {\bibfnamefont {R.}~\bibnamefont {Riedel}},\ }\href@noop {} {\bibfield
  {journal} {\bibinfo  {journal} {Phys. Status Solidi RRL}\ }\textbf {\bibinfo
  {volume} {5}},\ \bibinfo {pages} {196} (\bibinfo {year} {2011})}\BibitemShut
  {NoStop}%
\bibitem [{\citenamefont {Arab}\ \emph {et~al.}(2016)\citenamefont {Arab},
  \citenamefont {Sahraoui}, \citenamefont {Haddadi}, \citenamefont
  {Bouhemadou},\ and\ \citenamefont {Louail}}]{Arab2015}%
  \BibitemOpen
  \bibfield  {author} {\bibinfo {author} {\bibfnamefont {F.}~\bibnamefont
  {Arab}}, \bibinfo {author} {\bibfnamefont {F.~A.}\ \bibnamefont {Sahraoui}},
  \bibinfo {author} {\bibfnamefont {K.}~\bibnamefont {Haddadi}}, \bibinfo
  {author} {\bibfnamefont {A.}~\bibnamefont {Bouhemadou}}, \ and\ \bibinfo
  {author} {\bibfnamefont {L.}~\bibnamefont {Louail}},\ }\href@noop {}
  {\bibfield  {journal} {\bibinfo  {journal} {Phase Transit.}\ }\textbf
  {\bibinfo {volume} {89}},\ \bibinfo {pages} {480} (\bibinfo {year}
  {2016})}\BibitemShut {NoStop}%
\bibitem [{\citenamefont {Heyd}\ \emph {et~al.}(2003)\citenamefont {Heyd},
  \citenamefont {Scuseria},\ and\ \citenamefont {Ernzerhof}}]{Heyd2003}%
  \BibitemOpen
  \bibfield  {author} {\bibinfo {author} {\bibfnamefont {J.}~\bibnamefont
  {Heyd}}, \bibinfo {author} {\bibfnamefont {G.}~\bibnamefont {Scuseria}}, \
  and\ \bibinfo {author} {\bibfnamefont {M.}~\bibnamefont {Ernzerhof}},\
  }\href@noop {} {\bibfield  {journal} {\bibinfo  {journal} {J. Chem. Phys.}\
  }\textbf {\bibinfo {volume} {118}},\ \bibinfo {pages} {8207} (\bibinfo {year}
  {2003})}\BibitemShut {NoStop}%
\bibitem [{\citenamefont {Heyd}\ \emph {et~al.}(2006)\citenamefont {Heyd},
  \citenamefont {Scuseria},\ and\ \citenamefont {Ernzerhof}}]{Heyd2006}%
  \BibitemOpen
  \bibfield  {author} {\bibinfo {author} {\bibfnamefont {J.}~\bibnamefont
  {Heyd}}, \bibinfo {author} {\bibfnamefont {G.}~\bibnamefont {Scuseria}}, \
  and\ \bibinfo {author} {\bibfnamefont {M.}~\bibnamefont {Ernzerhof}},\
  }\href@noop {} {\bibfield  {journal} {\bibinfo  {journal} {J. Chem. Phys.}\
  }\textbf {\bibinfo {volume} {124}},\ \bibinfo {pages} {219906} (\bibinfo
  {year} {2006})}\BibitemShut {NoStop}%
\bibitem [{\citenamefont {Huang}\ \emph {et~al.}(2001)\citenamefont {Huang},
  \citenamefont {Tang},\ and\ \citenamefont {Lee}}]{Huang}%
  \BibitemOpen
  \bibfield  {author} {\bibinfo {author} {\bibfnamefont {J.~Y.}\ \bibnamefont
  {Huang}}, \bibinfo {author} {\bibfnamefont {L.-C.}\ \bibnamefont {Tang}}, \
  and\ \bibinfo {author} {\bibfnamefont {M.~H.}\ \bibnamefont {Lee}},\
  }\href@noop {} {\bibfield  {journal} {\bibinfo  {journal} {J. Phys.: Condens.
  Mater.}\ }\textbf {\bibinfo {volume} {13}},\ \bibinfo {pages} {10417}
  (\bibinfo {year} {2001})}\BibitemShut {NoStop}%
\bibitem [{\citenamefont {Fang}\ \emph {et~al.}(1999)\citenamefont {Fang},
  \citenamefont {de~Groot}, \citenamefont {Bruls}, \citenamefont {Hintzen},\
  and\ \citenamefont {de~With}}]{Fang1}%
  \BibitemOpen
  \bibfield  {author} {\bibinfo {author} {\bibfnamefont {C.~M.}\ \bibnamefont
  {Fang}}, \bibinfo {author} {\bibfnamefont {R.~A.}\ \bibnamefont {de~Groot}},
  \bibinfo {author} {\bibfnamefont {R.~J.}\ \bibnamefont {Bruls}}, \bibinfo
  {author} {\bibfnamefont {H.~T.}\ \bibnamefont {Hintzen}}, \ and\ \bibinfo
  {author} {\bibfnamefont {G.}~\bibnamefont {de~With}},\ }\href@noop {}
  {\bibfield  {journal} {\bibinfo  {journal} {J. Phys.: Condens. Mater.}\
  }\textbf {\bibinfo {volume} {11}},\ \bibinfo {pages} {4833} (\bibinfo {year}
  {1999})}\BibitemShut {NoStop}%
\bibitem [{\citenamefont {Bl{\"o}chl}(1994)}]{Blochl}%
  \BibitemOpen
  \bibfield  {author} {\bibinfo {author} {\bibfnamefont {P.~E.}\ \bibnamefont
  {Bl{\"o}chl}},\ }\href@noop {} {\bibfield  {journal} {\bibinfo  {journal}
  {Phys. Rev. B}\ }\textbf {\bibinfo {volume} {50}},\ \bibinfo {pages} {17953}
  (\bibinfo {year} {1994})}\BibitemShut {NoStop}%
\bibitem [{\citenamefont {Kresse}\ and\ \citenamefont
  {Furthm{\"u}ller}(1996)}]{KresseandFurth}%
  \BibitemOpen
  \bibfield  {author} {\bibinfo {author} {\bibfnamefont {G.}~\bibnamefont
  {Kresse}}\ and\ \bibinfo {author} {\bibfnamefont {J.}~\bibnamefont
  {Furthm{\"u}ller}},\ }\href@noop {} {\bibfield  {journal} {\bibinfo
  {journal} {Phys. Rev. B}\ }\textbf {\bibinfo {volume} {54}},\ \bibinfo
  {pages} {11169} (\bibinfo {year} {1996})}\BibitemShut {NoStop}%
\bibitem [{\citenamefont {Kresse}\ and\ \citenamefont
  {Joubert}(1999)}]{KresseandJoubert}%
  \BibitemOpen
  \bibfield  {author} {\bibinfo {author} {\bibfnamefont {G.}~\bibnamefont
  {Kresse}}\ and\ \bibinfo {author} {\bibfnamefont {D.}~\bibnamefont
  {Joubert}},\ }\href@noop {} {\bibfield  {journal} {\bibinfo  {journal} {Phys.
  Rev. B}\ }\textbf {\bibinfo {volume} {59}},\ \bibinfo {pages} {1758}
  (\bibinfo {year} {1999})}\BibitemShut {NoStop}%
\bibitem [{\citenamefont {Perdew}\ \emph {et~al.}(1996)\citenamefont {Perdew},
  \citenamefont {Burke},\ and\ \citenamefont {Ernzerhof}}]{PBE}%
  \BibitemOpen
  \bibfield  {author} {\bibinfo {author} {\bibfnamefont {J.~P.}\ \bibnamefont
  {Perdew}}, \bibinfo {author} {\bibfnamefont {K.}~\bibnamefont {Burke}}, \
  and\ \bibinfo {author} {\bibfnamefont {M.}~\bibnamefont {Ernzerhof}},\
  }\href@noop {} {\bibfield  {journal} {\bibinfo  {journal} {Phys. Rev. Lett.}\
  }\textbf {\bibinfo {volume} {77}},\ \bibinfo {pages} {3865} (\bibinfo {year}
  {1996})}\BibitemShut {NoStop}%
\bibitem [{\citenamefont {Perdew}\ \emph {et~al.}(2008)\citenamefont {Perdew},
  \citenamefont {Ruzsinszky}, \citenamefont {Csonka}, \citenamefont {Vydrov},
  \citenamefont {Scuseria}, \citenamefont {Constantin}, \citenamefont {Zhou},\
  and\ \citenamefont {Burke}}]{PBEsol}%
  \BibitemOpen
  \bibfield  {author} {\bibinfo {author} {\bibfnamefont {J.~P.}\ \bibnamefont
  {Perdew}}, \bibinfo {author} {\bibfnamefont {A.}~\bibnamefont {Ruzsinszky}},
  \bibinfo {author} {\bibfnamefont {G.~I.}\ \bibnamefont {Csonka}}, \bibinfo
  {author} {\bibfnamefont {O.~A.}\ \bibnamefont {Vydrov}}, \bibinfo {author}
  {\bibfnamefont {G.~E.}\ \bibnamefont {Scuseria}}, \bibinfo {author}
  {\bibfnamefont {L.~A.}\ \bibnamefont {Constantin}}, \bibinfo {author}
  {\bibfnamefont {X.}~\bibnamefont {Zhou}}, \ and\ \bibinfo {author}
  {\bibfnamefont {K.}~\bibnamefont {Burke}},\ }\href@noop {} {\bibfield
  {journal} {\bibinfo  {journal} {Phys. Rev. Lett.}\ }\textbf {\bibinfo
  {volume} {100}},\ \bibinfo {pages} {136406} (\bibinfo {year}
  {2008})}\BibitemShut {NoStop}%
\bibitem [{\citenamefont {Paier}\ \emph {et~al.}(2005)\citenamefont {Paier},
  \citenamefont {Hirschl}, \citenamefont {Marsman},\ and\ \citenamefont
  {Kresse}}]{Paier2005}%
  \BibitemOpen
  \bibfield  {author} {\bibinfo {author} {\bibfnamefont {J.}~\bibnamefont
  {Paier}}, \bibinfo {author} {\bibfnamefont {R.}~\bibnamefont {Hirschl}},
  \bibinfo {author} {\bibfnamefont {M.}~\bibnamefont {Marsman}}, \ and\
  \bibinfo {author} {\bibfnamefont {G.}~\bibnamefont {Kresse}},\ }\href@noop {}
  {\bibfield  {journal} {\bibinfo  {journal} {J. Chem. Phys.}\ }\textbf
  {\bibinfo {volume} {122}},\ \bibinfo {pages} {234102} (\bibinfo {year}
  {2005})}\BibitemShut {NoStop}%
\bibitem [{\citenamefont {Matsushita}\ \emph {et~al.}(2011)\citenamefont
  {Matsushita}, \citenamefont {Nakamura},\ and\ \citenamefont
  {Oshiyama}}]{Matsushita2011}%
  \BibitemOpen
  \bibfield  {author} {\bibinfo {author} {\bibfnamefont {Y.~I.}\ \bibnamefont
  {Matsushita}}, \bibinfo {author} {\bibfnamefont {K.}~\bibnamefont
  {Nakamura}}, \ and\ \bibinfo {author} {\bibfnamefont {A.}~\bibnamefont
  {Oshiyama}},\ }\href@noop {} {\bibfield  {journal} {\bibinfo  {journal}
  {Phys. Rev. B}\ }\textbf {\bibinfo {volume} {84}},\ \bibinfo {pages} {075205}
  (\bibinfo {year} {2011})}\BibitemShut {NoStop}%
\bibitem [{\citenamefont {R{\aa}sander}\ and\ \citenamefont
  {Moram}(2015)}]{Rasander2015}%
  \BibitemOpen
  \bibfield  {author} {\bibinfo {author} {\bibfnamefont {M.}~\bibnamefont
  {R{\aa}sander}}\ and\ \bibinfo {author} {\bibfnamefont {M.~A.}\ \bibnamefont
  {Moram}},\ }\href@noop {} {\bibfield  {journal} {\bibinfo  {journal} {J.
  Chem. Phys.}\ }\textbf {\bibinfo {volume} {143}},\ \bibinfo {pages} {144104}
  (\bibinfo {year} {2015})}\BibitemShut {NoStop}%
\bibitem [{\citenamefont {Setyawan}\ and\ \citenamefont
  {Curtarolo}(2010)}]{Setyawan2010}%
  \BibitemOpen
  \bibfield  {author} {\bibinfo {author} {\bibfnamefont {W.}~\bibnamefont
  {Setyawan}}\ and\ \bibinfo {author} {\bibfnamefont {S.}~\bibnamefont
  {Curtarolo}},\ }\href@noop {} {\bibfield  {journal} {\bibinfo  {journal}
  {Comp. Mater. Sci.}\ }\textbf {\bibinfo {volume} {49}},\ \bibinfo {pages}
  {299} (\bibinfo {year} {2010})}\BibitemShut {NoStop}%
\end{thebibliography}%

\end{document}